\documentstyle[11pt,aasms4]{article}

\received{}
\accepted{}

\lefthead{Dupac {\it{et al.}}}
\righthead{Submillimeter mapping and analysis of cold dust condensations
in the Orion M42 star forming complex}

\begin{document}

\def\etal{{\it et al.} }
\def\araa{{\it Ann.\ Rev.\ Astron.\ Ap.}}
\def\aplet{{\it Ap.\ Letters}}
\def\aj{{\it Astron.\ J.}}
\def\apj{{\it ApJ}}
\def\apjl{{\it ApJ\ (Lett.)}}
\def\apjs{{\it ApJ\ Suppl.}}
\def\aas{{\it Astron.\ Astrophys.\ Suppl.}}
\def\aa{{\it A\&A}}
\def\mnras{{\it MNRAS}}
\def\nature{{\it Nature}}
\def\pre{{\it Preprint}}
\def\aph{{\it Astro-ph}}
\def\expas{{\it Experimental Astron.}}
\def\inpress{{\it in press}}
\def\inprep{{\it in preparation}}
\def\submit{{\it submitted}}

\def\ap{$\approx$ }
\def\mjysr{MJy/sr }
\def\inu{{I_{\nu}}}
\def\inufit{I_{\nu fit}}
\def\fnu{{F_{\nu}}}
\def\bnu{{B_{\nu}}}
\def\msol{{M$_{\odot}$}}
\def\mic{{{\mu}m}}
\def\cm2{$cm^{-2}$}

\title{Submillimeter mapping and analysis of cold dust condensations
in the Orion M42 star forming complex} 

\author{X. Dupac \altaffilmark{1}, M. Giard\\
Centre d'\'Etude Spatiale des Rayonnements (CESR)\\ 
9 av. du colonel Roche, BP4346, F-31028 Toulouse cedex 4, France}

\author{J.-P. Bernard, J.-M. Lamarre\\ 
Institut d'Astrophysique Spatiale (IAS)\\
Campus d'Orsay B\^at. 121 \\
15, rue Cl\'emenceau, F-91405 Orsay cedex, France}

\author{C. M\'eny\\
CESR}

\author{F. Pajot\\
IAS}

\author{I. Ristorcelli, G. Serra \altaffilmark{2} \\
CESR}

\author{J.-P. Torre \\
Service d'A\'eronomie du CNRS \\
BP3, F-91371 Verri\`eres-le-Buisson cedex, France}

\altaffiltext{1}{{\it Send offprint requests to} dupac@cesr.fr}
\altaffiltext{2}{Guy Serra, principal investigator of the ProNaOS project,
  passed away on August the $15^{th}$ 2000.}

\begin{abstract}

We present here the continuum submillimeter maps of the molecular cloud
around the M42 Nebula in the Orion region. 
These have been obtained in four wavelength
bands (200, 260, 360 and 580 $\mic$) with the ProNaOS two meter
balloon-borne telescope. The area covered is 7 parsecs wide (50 arcmin
at a distance of 470 pc) with a spatial resolution of about 0.4 parsec.
Thanks to the high sensitivity to faint surface brightness gradients,
we have found several cold condensations with temperatures
ranging from 12 to 17 K, within 3 parsecs of the dense ridge. The statistical analysis of the temperature and spectral index spatial
distribution shows an evidence of an inverse correlation
between these two parameters.
Being invisible in the IRAS 100 $\mic$ survey, some cold clouds
are likely to be the seeds for future star formation activity
going on in the complex. We estimate their masses and we show that two of them
have masses higher than their Jeans masses, and may be gravitationally
unstable.

\end{abstract}


\section{Introduction\label{intro}}

At about 470 pc, the Orion Nebula is one of the nearest massive star formation
regions.
It is part of the giant molecular complex of Orion, that has been
revealed by the $^{12}$CO maps of Kutner \etal (1977).
The CO map of Maddalena \etal (1986) has shown the global distribution of gas
in the Orion-Monoceros complex.
This dense region of the
interstellar medium, extending over approximatively 30 degrees from southeast
to northwest, is composed of several distinct areas.
The most northern part is the gaseous region distributed on a circle around
$\lambda$ Orionis. To the south, the Orion Giant Molecular Clouds A and B extend
over 15 degrees from north to south.
To the southeast are located Monoceros and the Southern Filament.
This set of giant molecular clouds is located 150 pc below the Galactic plane,
and might have been formed by the collision of a cloud falling from the
southern Galactic hemisphere onto the galactic disk (\cite{franco88}).
These molecular clouds are associated with large HII regions, the Orion Ia,
Ib, Ic OB star associations, located on a 15 degree long ridge on the side
away from the Galactic plane.
The Orion Nebula is part of the Orion A giant molecular cloud, and its best studied region.
Behind this H II region is
the OMC-1 molecular cloud, which corresponds to a CO emission peak. This cloud
is heated by the ultraviolet radiation of the OB stars of the Trapezium
cluster (see for example \cite{hillenbrand97}).
One can distinguish inside of OMC-1, two intense
infrared sources: the ponctual object of Becklin-Neugebauer (\cite{becklin67})
and, 10
arcseconds to the south, the Kleinmann-Low Nebula (\cite{kleinmann67}) that has an angular
diameter of about 30 arcseconds. It is thought that BN is a young massive star (25 \msol) with an
important mass loss.
The first far infrared observation towards M42 has been performed by Low \& Aumann (1970) in the spectral range from 30
$\mic$ to 1000 $\mic$. Then Harper (1974) mapped M42 at 90 $\mic$ and showed
that the infrared emission originates from a more extended region.
This area has been since intensively studied, from optical radiations to radio wavelengths (see for example
the review of \cite{genzel89}).
North from OMC-1, one can clearly see an extended "integral-shaped" filament
(ISF hereafter) with two denser
regions called OMC-2 and OMC-3 (see \cite{bally87}).
These cores have been studied both
in molecular lines (recent studies of \cite{dutrey93}, \cite{castets95}, \cite{nagahama98})
and continuum (recent works of \cite{chini97}, \cite{lis98},
\cite{johnstone99}).
The 1300 $\mic$ maps of OMC-2 and OMC-3 of Chini \etal (1997) have shown the
precise filamentary structure of these regions, studying several sources inside.
On the dense ridge, the three cores OMC-1, OMC-2, OMC-3 have active star
formation and seem to have been formed at roughly the same time. However there
are important differences between them. OMC-1 has massive star formation and
strong interactions between the gas and the infrared sources, increasing the
gas temperature to $\approx$ 70 K, whereas OMC-2 is less luminous, contains
low mass young stars and less energetic outflows than OMC-1. OMC-3 is thought
to have a mass comparable to OMC-2 ($\approx$ 100 \msol), but has less
energetic outflows and lower gas temperature, suggesting that OMC-3 is less
evolved than OMC-2, itself being less evolved than OMC-1 (see
\cite{castets95}).
The recent works in submillimeter continuum (\cite{lis98}, \cite{johnstone99}) have high
angular resolution, and thus have clearly shown the complex structure of the
Orion Integral-Shaped Filament and discovered small condensations, as also did
the $^{13}CO$ work of Nagahama \etal (1998) for the whole Orion A cloud.
However the continuum ground-based observations do not have the sensitivity to map the
weak emission from the dust away from the dense ridge.
A recent far-infrared balloon-borne observation of Orion (\cite{mookerjea00})
has derived the temperature distribution in Orion, the coldest source they
found being 15 K.

We have observed this region with the submillimeter balloon-borne telescope ProNaOS, in order to better understand
the variations of temperature and spectral index of the dust, and to study the
cold phase of the dense interstellar medium.
ProNaOS provides simultaneous measurements within four bands covering the wavelength range
200 $\mic$ - 800 $\mic$, allowing to constrain both the temperature
and the spectral index of the dust, in order to derive column densities and
masses.
The high sensitivity per beam allows us to study the extended emission of the
dust, in and a few parsecs away from the dense ridge.
A first ProNaOS mapping of OMC-1 and its surroundings has been analyzed by
Ristorcelli \etal (1998). They discovered a cold (12.5 K) condensation and
showed for four sources the variability of temperature and spectral index
in the Orion Nebula complex.
We present in this paper a detailed analysis of the submillimeter maps
obtained towards Orion with the second flight of ProNaOS in september 1996, at Fort-Sumner, New Mexico.
We have observed a much larger region than in Ristorcelli \etal including the integral-shaped
filament and other cold clouds, two being seen for the first time in continuum
emission.
Section 2 deals with
the observation performed, section 3 describes the maps obtained, and section 4
presents a detailed analysis of these maps.

\section{Instruments and observations\label{obs}}

\subsection{Observations with ProNaOS}

ProNaOS (PROgramme NAtional d'Observations Submillim\'etriques) is a French
balloon-borne submillimeter experiment, with a 2 m diameter telescope (\cite{buisson90}). The main characteristics of the instrument are given in Table 1.
The
focal plane instrument SPM (Syst\`eme Photom\'etrique Multibande, see \cite{lamarre94}) is
composed of a wobbling mirror, providing a beam switching on the sky with an
amplitude of about 6' at 19.5 Hz, and four bolometers cooled at 0.3 K. They
measure the submillimeter flux in the spectral ranges 180-240 $\mic$, 240-340
$\mic$, 340-540 $\mic$ and 540-1200 $\mic$, with sensitivity to low
brightness gradients of about 1 \mjysr in band 4.
Thanks to an arrangement with dicroic filters, the measurement is performed
simultaneously for all four bands on the same sky pixel.
An off-axis star tracker and a 3-axis gyroscope allow to control the telescope
pointing.
Before a flight, the telescope and the whole payload are aligned and
controlled.
Two internal black bodies are calibrated with an absolute
reference.
They provide in-flight calibrations with an accuracy of about 5 \%
(relative), which has been checked against Saturn.
The measurements on Saturn allow us to derive the
beam shape up to a radius of six arcminutes from
the axis.
We use the integral of these beams to compare
the Saturn measurement to the ground based calibration
which is performed on extended blackbodies filling
the beam.
Regarding the ratio of the main central 
beam to the total beam, the beam efficiencies are from 0.78 in band 1 to 0.96
in band 4.
This can be expressed as a mean wave-front error of about 15 $\mic$.
More details about the instrument can be found in Ristorcelli \etal (1998).

The observing procedure is an altazimut scanning of the beam on the
sky. The time-ordered data suffer thus from two main artifacts: the measured
flux is differential, and the rotations of the telescope around the direction
of observation distort the mapping pattern from a rectangular grid.

\begin{table}
\caption{ProNaOS instrument characteristics}
\begin{center}
\begin{tabular}{|l|cccc|}
\hline
Primary mirror diameter & \multicolumn{2}{c}{2045 mm, f/10}&  &  \\
Wobbling mirror & \multicolumn{2}{c}{f=19.5 Hz, $\delta _{cross-elevation}=6'$}&  & \\
Pointing accuracy & \multicolumn{2}{c}{$20"$ absolute, $5"$ relative}& & \\
Band& 1 & 2 & 3 & 4 \\
($\mic$) &  $180-240$ & $240-340$ & $340-540$ & $540-1200$ \\
Beam diameter (arcmin)& 2 &  2  &  2.5  & 3.5 \\
NEB ($MJy sr^{-1}Hz^{-1/2}$)& 22 & 26 & 8.5 & 5.3 \\
\hline
\end{tabular}
\end{center}
\end{table}

\subsection{Data processing: a new method for ProNaOS data}

The first processing applied to the data is correction from map
distortion, taking into account the pointing of the telescope, including fine
pointing errors due to the swinging of the gondola.
Then usually the
deconvolution method applied to ProNaOS data was EKH-like (\cite{emerson79})
with a scan by scan filtering in Fourier space (see \cite{sales91}).
We have developed another method, based on direct linear inversion on the
whole map, using a Wiener matrix.
The Wiener filter (\cite{wiener49}) is the linear method that minimizes the
reconstruction error.
The restored sky map is computed as a linear expression of the data:

$x_{0} = W y$

$x_{0}$ being the reconstructed sky vector (i.e. the map), W the inversion
matrix and y the signal vector. Assuming that the instrument is linear, the
observation is:

$y = A x + n$

where x is the true sky vector, A the observation matrix and n the noise vector.
For W we use a Wiener matrix (see \cite{tegmark97}):

$W = [S^{-1} + A^{t}N^{-1}A]^{-1} A^{t}N^{-1} $

where S is the sky covariance matrix and N the noise
covariance matrix.
The convolution matrix A (point-spread function) takes into account the size
and profile of the two beams, and the observing mode.
In the case of the Orion region presented here, we have gathered two
observation sequences.
The first one is centered 10' north from BN/KL, and covers 50' by 40'.
The
second is offset to the west to map the edge of the molecular cloud, and
covers about 30' by 20'.
In such data, one of the main problems encountered to properly reconstruct the
maps, is the great contrast between the intense areas like OMC-1 and the weak
areas, which are not less interesting.
We have processed separatly the second sequence, in order to decrease the
contrast in these data and reconstruct precisely the intensities of the second sequence.
The size of A is about 2000*3000, which allows us to use a
classical bi-conjugate gradient algorithm to invert the main term in W.
The correlation matrices are estimated from the data, and can be refined with
iterations and tests.

We show in Fig. 1 the map of the first sequence obtained with the EKH-like
method, to be compared in Fig. 2 to the optimal maps obtained with the Wiener
inversion method.
Negative artifacts are clearly visible along the scans of the most
intense area (along the elevation of OMC-1) with the EKH-like method, whereas
they are significantly removed with the Wiener inversion method.

\begin{figure}[ht]
\caption[]{200 $\mic$ map obtained with the EKH-like method (first observation
  sequence only).
The color scale is logarithmic, and displays the positive reconstructed flux
until 0.5 in log, then the light blue and the deep blue display the negative
noise features, clearly visible along the elevation of OMC-1.
The black lines in the color bar show the contour levels.
}
\end{figure}

\subsection{Observations with DiaBolo}

Observations of Cloud 2 have been performed with the DiaBolo photometer at the 
focus of the IRAM 30 meter telescope in december 6/7 1996. DiaBolo is a two
channel (1200 $\mic$ and 2100 $\mic$) photometer
which uses 100 mK bolometers. A detailed description of the instrumental setup can be found 
in Beno\^\i t {\it{et al.}} (2000). The configuration for the observation of Cloud 2 used one bolometer per wavelength channel with a beam size of about 30 arcseconds. 
We used the wobbling secondary mirror to subtract the background emission from the 
sky and the telescope. The wobbling amplitude was set to 120 arcseconds for a frequency 
of 0.8 Hz. We mapped a 6 by 6 arcminute area centered on the cloud by scanning the 
telescope beam in azimut and elevation. The scanning speed was 6 arcsec per second, 
and the step between two lines was 15 arcseconds.
The maps were deglitched by filtering the signal of each
scan using a temporal filter.
After base line subtraction and rebinning of each
scan, the dual beam maps were deconvolved from the beam switching using
home-made IDL
routines which follow the EKH method (\cite{emerson79}).

\section{Results\label{res}}

We present in Fig. 2 the images obtained in the two extreme photometric bands of
ProNaOS-SPM.
The noise level is about 4 \mjysr rms in band 1 and 0.8 \mjysr in band 4.
However, due to the calibration uncertainty, the flux accuracy is not better
than 5 \% (1 $\sigma$) relative between bands (10-20 \% absolute).

\begin{figure}[ht]
\caption[]{ProNaOS maps in band 1 (200 $\mic$, up) and band 4 (580 $\mic$,
  bottom).
For the band 1 image, we have merged the two sequences, reconstructed
separatly, into one image.
The color scale is logarithmic, and displays the positive reconstructed flux
until -1.5 in log, then the deeper blue and purple colors display the negative
noise features.
The noise level is about 4 \mjysr rms in band 1 and 0.8 \mjysr in band 4.
However, due to the calibration uncertainty, the flux accuracy is not better
than 5 \% (1 $\sigma$) relative between bands (10-20 \% absolute).
The black lines in the color bar show the contour levels.
The black box drawn is the area mapped by Ristorcelli \etal (1998).
}
\end{figure}

The brightest area observed is the molecular cloud OMC-1,
which appears on our maps as a very intense region of thermal emission of the
dust. Particularly in the area of
BN/KL, the emission is very strong (it is the maximum of the submillimeter
emission in this region): it reaches 49000 MJy/sr in band 1 (200 $\mic$).

North from this intense central core is clearly visible a large molecular cloud
extending over about 3 pc, called the integral-shaped filament, which presents an
average spectral intensity at 200 $\mic$ of about 2300 MJy/sr.

One can see west from the central core a weaker condensation that we call Cloud 1, which has an intensity
of 570 MJy/sr in band 1. A weaker emission can be seen around this cloud, which
is linked to the Cloud 2 northwards.

The Clouds 2, 3, 4 can be seen on the west of the maps in Fig. 2, from
east to west. The Cloud 2 is a weak intensity condensation discovered during the first flight of ProNaOS
(\cite{ristorcelli98}). Its 200 $\mic$ intensity is 47
MJy/sr, but its maximum is closer to band 2 (260 $\mic$), in which the spectral
intensity reaches 50 MJy/sr.

The Clouds 3 and 4 are two extremely weak condensations.
The 260 $\mic$ intensity of Cloud 3 is 32 MJy/sr, and the 200 $\mic$
intensity of Cloud 4 is 35 MJy/sr.
These cold clouds, as well as Cloud 2, cannot be seen with IRAS 100 $\mic$
maps of the same region.

\section{Analysis\label{ana}}

\subsection{Dust temperatures and spectral indexes\label{fit}}

\subsubsection{Derivation}

To compare the emission obtained in the four submillimeter bands, and
particularly to deduce emission spectra
of the sources identified, it is necessary to degrade the angular resolution
in all channels to the same beam.
This is why we smooth the images in bands 1, 2, 3
with an adequate profile in order to obtain for each band the resolution of the
fourth band (3.5'). The corresponding beams have been determined from the
observation performed on the planet Saturn, considered as a reference point source.

The ProNaOS integrated fluxes in the four bands are given in Table 2. The integrated
fluxes from other data (IRAS: http://www.ipac.caltech.edu/ipac/iras/iras.html,
\cite{harper74}, \cite{chini84}, \cite{mezger90}, DiaBolo: \cite{benoit00}) are given in
Table 3.

\begin{table}
\caption[]{\label{table: fluxes}
Integrated fluxes for ProNaOS data. The diameters of the regions are given in arcminutes.}
\begin{flushleft}
\begin{tabular}{lllllll}

\hline
 & $\alpha_{1950}$ & $\delta_{1950}$ & F$_{\nu}$(Jy) & F$_{\nu}$(Jy) & F$_{\nu}$(Jy) & F$_{\nu}$(Jy) \\
 & (h,min,sec) & ($^o$,') & 200 $\mic$ & 260 $\mic$ & 360 $\mic$ & 580 $\mic$ \\
\hline

OMC-1(3.6') & 5h 32min 38& -5$^o$ 25'& 28500 $\pm $ 1400 & 14120 $\pm $ 710 & 6010 $\pm $ 300 & 1413 $\pm $ 71\\
\hline

ISF & & & & & &\\
(south) 5.4' & 5h 32min 52& -5$^o$ 13'& 5130 $\pm $ 260 & 3110 $\pm $ 160 & 1489 $\pm $ 74 & 404 $\pm $ 20\\
\hline

ISF & & & & & &\\
(north) 4.2'  & 5h 32min 38& -4$^o$ 58'& 2490 $\pm $ 120 & 1449 $\pm $ 72 & 699 $\pm $ 35 & 182.9 $\pm $ 9.1\\
\hline

Cloud1 (3.5') & 5h 31min 30& -5$^o$ 25'& 461 $\pm $ 23 & 308 $\pm $ 15 & 209 $\pm $ 10 & 51.0 $\pm $ 2.6\\
\hline

Cloud2 (3.5') & 5h 31min 16& -5$^o$ 08'& 37.9 $\pm $ 1.9 & 41.0 $\pm $ 2.0 & 27.0 $\pm $ 1.3 & 9.95 $\pm $ 0.50\\
\hline

Cloud3 (3.5') & 5h 30min 15& -5$^o$ 01'& 25.0 $\pm $ 1.3 & 25.8 $\pm $ 1.3 & 15.26 $\pm $ 0.76 & 5.82 $\pm $ 0.29\\
\hline

Cloud4 (5.2') & 5h 29min 11& -4$^o$ 57'& 63.2 $\pm $ 3.2 & 59.5 $\pm $ 3.0 & 21.5 $\pm $ 1.1 & 12.54 $\pm $ 0.63\\
\hline

\end{tabular}
\end{flushleft}
\end{table}

\begin{table}
\caption[]{\label{table: fluxes2}
Integrated fluxes from other data. The
approximative diameters of the regions are given in arcminutes.}
\begin{flushleft}
\begin{tabular}{lllllll}
\hline

& F$_{\nu}$(Jy) & F$_{\nu}$(Jy) & F$_{\nu}$(Jy) & F$_{\nu}$(Jy) &
F$_{\nu}$(Jy) & F$_{\nu}$(Jy) \\
& \tiny{90 $\mic$ Harper} & \tiny{100 $\mic$ IRAS} & \tiny{1000 $\mic$ Chini} & \tiny{1200 $\mic$ DiaBolo}
& \tiny{1300 $\mic$ Mezger} & \tiny{2100 $\mic$ DiaBolo} \\
\hline

OMC-1(3.6') & \small{258000 $\pm $ 77000} & --- & \small{480 $\pm$ 140} & --- &
\small{187 $\pm$ 56} & --- \\
\hline

ISF &&&&&& \\
(south) & --- & \small{10600 $\pm $ 2600} & --- & --- & --- & --- \\
\hline

ISF &&&&&& \\
(north) & --- & \small{4000 $\pm $ 1000} & --- & --- & --- & --- \\
\hline

Cloud2 (3.5') & --- & --- & --- & \small{0.77 $\pm $ 0.13} & --- & \small{0.120 $\pm $ 0.030} \\
\hline

\end{tabular}
\end{flushleft}
\end{table}

We model the emission of the grains with a modified blackbody law:

$\inufit(\lambda,T,n)=C . \bnu(\lambda,T) . \lambda^{-\beta}$

where $\lambda$ is the wavelength, C a constant, T the temperature of the grains, $\beta$ the spectral
index and $\bnu$ the Planck function.

The three parameters C, T and $\beta$, are adjusted with a least square fit.
Assuming optically thin emission (as did \cite{ristorcelli98}), we can derive
the optical depth $\tau_\nu={I_\nu\over{B_\nu}}=C . \lambda^{-\beta}$.

For some regions: Integral-Shaped Filament (south) and ISF (north), we include the
IRAS
100 $\mic$ data in the fit. For the other regions, we cannot use these data, because they are either saturated (OMC-1), or drowned in the
halo of OMC-1 (Clouds 1,2,3,4).
For the fit of OMC-1, we included the 90 $\mic$ data of Harper
(1974), the 1000 $\mic$ data of Chini \etal (1984), and the 1300 $\mic$ data
of Mezger \etal (1990). These data have been averaged over the same beam diameter
as ProNaOS, i.e. 3.5'.

For Cloud 2, we added the data obtained with the millimeter instrument DiaBolo
at 1200 $\mic$ and 2100 $\mic$.
Three 
maps were repeated on Cloud 2, for a total observing time of about 1 hour. 
The values given in Table 3 have been obtained by integrating in 
a 3.5 arcmin circle centered on the cloud. 

We give spectra of some regions in Fig. 3.
The best parameters for each region selected are presented in Table 4.

\begin{figure}[ht]
\epsfbox{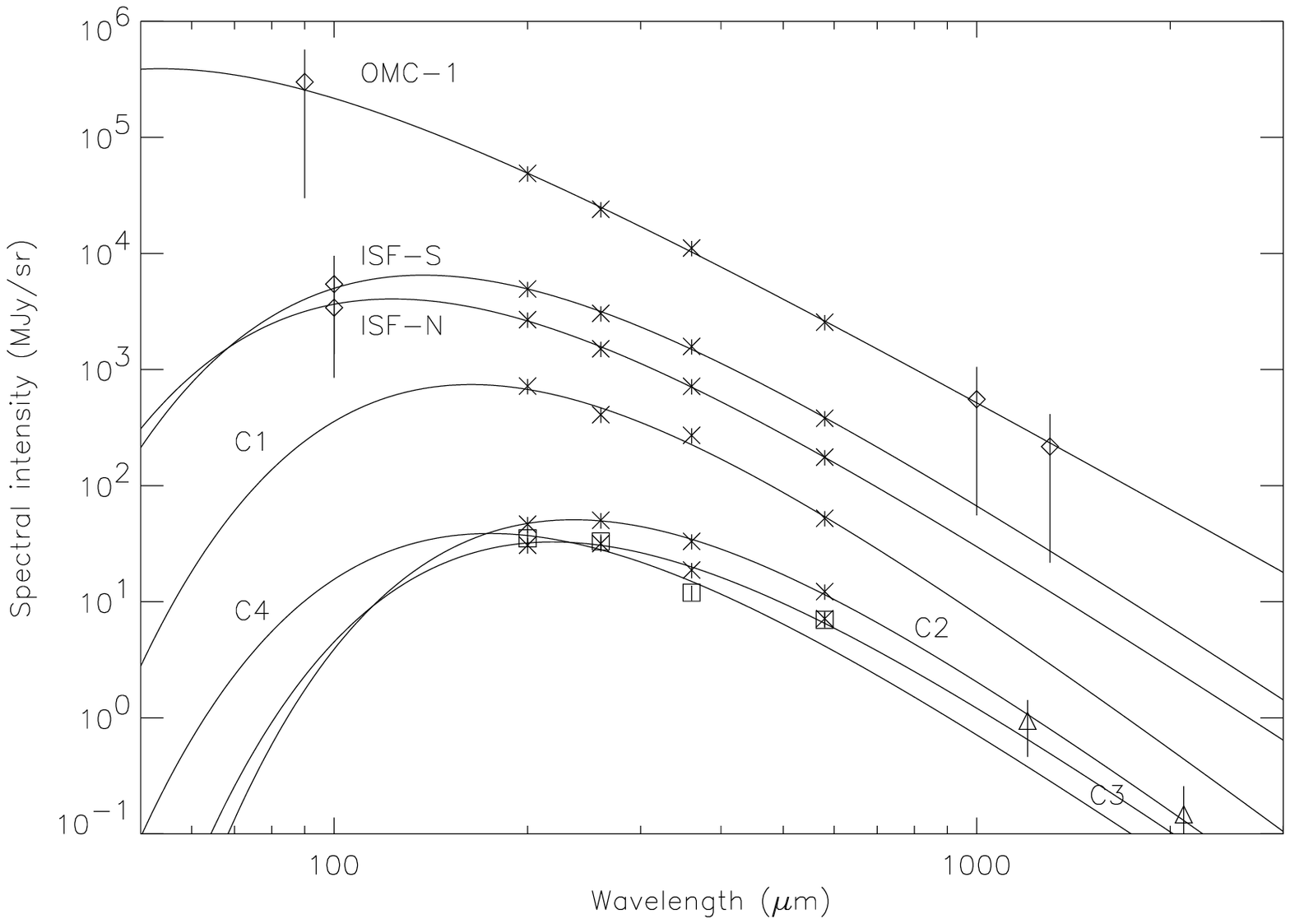}
\caption[]{Spectra with 3 $\sigma$ error bars.
The crosses stand for the ProNaOS points, the triangles for the DiaBolo points
(Cloud 2), and the diamonds for the other data.
For clarity, the Cloud 4 ProNaOS points appear as squares.
The ProNaOS error bars are the intercalibration errors.
The drawn lines are the result of the fits (modified black body).}
\end{figure}

\begin{table}
\caption[]{\label{table: results}
Temperature and spectral index from the fit of the spectra, optical
depth. The error bars are given for 68 \% confidence interval. The
approximative diameters of the regions are given in arcminutes.}

\begin{flushleft}
\begin{tabular}{lllll}
\hline
   & T (K) & $\beta$ & ${\tau_{\nu}}\over{10^{-3}}$ & ${\tau_{\nu}}\over{10^{-3}}$\\
   & & & 200 $\mic$ & 580 $\mic$\\

\hline

OMC-1(3.6') & 66.1 $^{+10.2} _{-9.4}$ &
1.13 $^{+0.06} _{-0.09}$ & 47 $\pm $ 6 & 10.4 $\pm $ 0.8\\
\hline

ISF & & & & \\
(south) 5.4' & 22.4 $^{+2.1} _{-2.0}$ & 1.71 $^{+0.12} _{-0.19}$ & 5.3 $\pm $ 1.0 & 1.3 $\pm $ 0.1\\
\hline

ISF & & & & \\
(north) 4.2'  & 25.2 $^{+2.5} _{-2.5}$ & 1.68 $^{+0.13}
_{-0.17}$ & 15 $\pm $ 3 & 1.9 $\pm $ 0.2\\
\hline

Cloud1 (3.5') & 17.0 $^{+3.4} _{-2.5}$ & 2.21 $^{+0.23}
_{-0.48}$ & 4.0 $\pm $ 3.2 & 0.76 $\pm $ 0.23\\
\hline

Cloud2 (3.5') & 11.8 $^{+0.6} _{-0.7}$ & 2.20
$^{+0.15} _{-0.18}$ & 4.2 $\pm $ 1.5 & 0.43 $\pm $ 0.07 \\ 
\hline

Cloud3 (3.5') & 13.3 $^{+2.6} _{-2.1}$ &
1.98 $^{+0.39} _{-0.79}$ & 1.4 $<4.2 (3 \sigma)$ & 0.19 $\pm $ 0.07\\
\hline

Cloud4 (5.2') & 16.9 $^{+7.3} _{-4.1}$ & 1.91 $^{+0.00}
_{-1.29}$ & 0.5 $<2.1 (3 \sigma)$ & 0.12 $\pm $ 0.06\\
\hline

\end{tabular}
\end{flushleft}
\end{table}

\subsubsection{Variations of the temperature and the spectral index}

The dust temperature is highly variable all around this
molecular complex: from {\bf 12 K} to {\bf 70 K}; and the spectral index changes much too:
from {\bf 1.1} to {\bf 2.2}, as was also found for the four sources analyzed
by Ristorcelli \etal (1998).
The clouds separated from the denser regions (Clouds 3,4) are cold
and have high indexes, but we can also see that some cold condensations (Cloud
1, 17 K, Cloud 2, 11.8 K) with high indexes can
be close to the central region OMC-1 (1.5 pc).
For the two sources in common (OMC-1 and Cloud 2), the parameters derived in
this paper are well consistent with Ristorcelli \etal
The spectral index derived for the OMC-1 cloud is 1.13 with quite small error
bars, which is due to the high number of data points we have in the
Rayleigh-Jeans tail of the dust emission, the temperature being around 70 K
(see \cite{ristorcelli98} for a precise discussion on this region).
For such a region, the submillimeter spectrum we observe could come from a
mixture of several dust temperatures.
It has been shown by Ristorcelli \etal (1998) that the cold dust column density
required to explain the fit temperature (83 K) with $\beta=2$ was 40 times more massive than the warm component.
They concluded that the spectral index of the dominant component had to be around 1.
The ISF extended emission is quite cold (22 K - south, 25 K - north) with
spectral indexes of about 1.7.
In this temperature range, the ProNaOS data points combined with the IRAS 100
$\mic$ point give quite small error bars.
We have simulated a temperature distribution with $\beta=2$, with a bell
function around 17 K having a fwhm of about 9 K.
This leads to fit the spectrum with a temperature of 21 K and a spectral index
of 1.8.
Thus the temperature and spectral index range we observe in the ISF could
partly be explained by a distribution of temperatures with a colder central
temperature, and a higher index.
The clouds 1 and 2 have the highest spectral index we found in this region
(2.2), Cloud 2 being the coldest (11.8 K), with error bars considerably
reduced, thanks to the DiaBolo millimeter data, with respect to the parameters given in Ristorcelli \etal
The Cloud 3 is another cold cloud (13.3 K) with a high index (1.98), however
the error bar on the spectral index is large.
The Cloud 4 has ill-defined parameters, but seems also to have comparable values
of the temperature and the spectral index.

Recently, high angular resolution submillimeter maps of the integral-shaped filament have
been obtained from the ground using the CSO at 350 microns (\cite{lis98}) and
the JCMT at 450 and 850 microns (\cite{johnstone99}).
However, these maps do not include the fainter regions
evidenced here (Clouds 1-4) nor the diffuse emission around the ISF that
we show in this article.
Comparing their maps to the ProNaOS ones, it is clear that the extended
emission of the dust around the brightest sources is missed by these
ground-based experiments.
Combining their 350 $\mic$ map and the IRAM 1.3 mm
map of
Chini \etal (1997), Lis \etal derived a map of the dust emissivity index over
the brightest region surrounding OMC-1, including Orion-S and the Orion bar.
Under the assumption of a single dust temperature (55 K) over this region, they
derived emissivity indexes in the range 1.3 to 2.7, the lower values being
observed toward the Orion bar (about 1.7) and IRC2 (1.8) while higher values
(typically 2.3) appear north of OMC-1. The index values they derived toward
the brightest region of the map (IRC2 and Orion-S) is in the range 1.8-2.0
when averaged within the beam size of our observations. Even if we take into
account that the temperature they assumed is too low compared to the one
we measure (66 K), these
values remain clearly higher than the dust emissivity index value we derived
for this region ($\beta$=1.13).
This discrepancy may originate
from their use of two distinct datasets with different systematic calibration
errors.

\subsection{Temperature - index inverse correlation\label{corr}}

We have computed the 68 \% (1 $\sigma$) confidence likelihood contours of the (T,$\beta$) values
for each cloud (see Fig. 4) -{\it N.B.}: the
likelihood is proportional to $e^{-\chi^{2}\over2}$.

\begin{figure}[ht]
\epsfbox{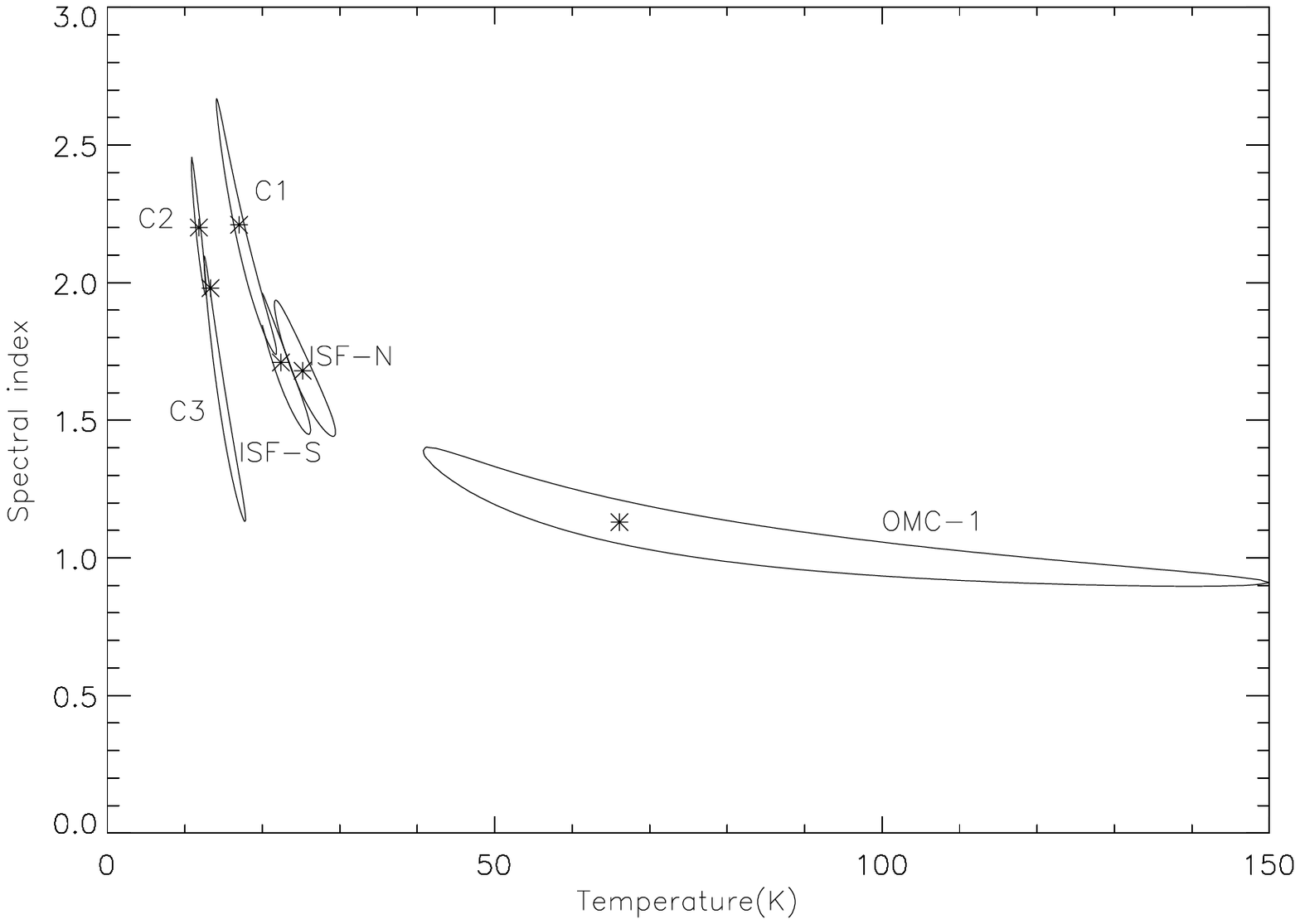}
\caption[]{1 $\sigma$ likelihood contours of six regions of M42.
The drawn crosses are the positions of the maximum likelihoods.}
\end{figure}

The correlation coefficient between T and $\beta$ has been computed globally for the
studied regions. We have therefore seven couples (T,$\beta$) which correlation we
analyse by:

$C={\sum{(T_{i}-\bar T) . (\beta_{i}-\bar \beta)}\over\sqrt{\sum{(T_{i}-\bar T)^{2}}
  . \sum{(\beta_{i}-\bar \beta)^{2}}}}$

A first analysis of the plot in Fig. 4 shows an inverse correlation between
the temperature and the spectral index, which means that the regions
of high index have the lowest temperature, and the regions of low index have
the highest temperature.
The correlation coefficient calculated is
{\bf -0.92}.
This correlation is, for a part, dominated by the warm area of OMC-1, but the correlation coefficient is still {\bf -0.83} when this area is not included.

One can see in Fig. 4 that the likelihood contours are banana-shaped, that is
mainly due for OMC-1 to the insensitivity to temperature variations in the
Rayleigh-Jeans tail, and for the cold clouds to the relatively bad sensitivity
to spectral index variations, as we do not have millimeter data for most of
them.
However the fit procedure itself could induce an amount of inverse correlation
between the temperature and the spectral index.
To prove that this correlation is for the most part not
due to the fit, we performed simulations. Repeated fits on noisy simulated data
with no intrinsic (T,$\beta$) correlation show that the fitting procedure itself induces a correlation coefficient lesser in
absolute than {\bf -0.4}. This artifact of the fit is thus insufficient to explain the
correlation found on real data.

Our conclusion is that this inverse correlation found between the temperature
and the spectral index has
to be an
intrinsic physical property of the grains.

Laboratory experiments (see
\cite{agladze96} and \cite{mennella98}) showed this effect for temperatures
down to 10 K, but it had never been systematically shown until now with
observations.
Agladze \etal (1996) measured absorption spectra of crystalline and amorphous
grains between 0.7 and 2.9 mm wavelength. They deduced an inverse correlation
between the power-law index $\beta$ and the temperature in the temperature range
10-25 K, and attributed it to two level tunnelling processes.
The measures of Agladze \etal are for us insufficient to justify
our observation in the submillimeter spectral range, because absorption can be
very different than in the millimeter range.
Mennella \etal (1998) measured the absorption coefficient of cosmic dust
analogue grains, crystalline and amorphous, between 20 $\mic$ and 2 mm
wavelength, in the temperature range 24-295 K. They deduced an inverse
correlation between T and $\beta$, and attributed it to two phonon difference processes.

\subsection{Column densities\label{model}}

\subsubsection{Derivation}

To calculate the opacity of the grains, the masses and densities of the
studied regions, we need to estimate the column density in each pixel of the
maps, i.e. the mass of gas and dust along the line of sight. For this we use
the dust 100 $\mic$ opacity from D\'esert \etal (1990). It allows us to
estimate the column density from the spectral intensity. We consider
only the thermal emission of the big grains, which dominate widely in this
spectral range, and we assume that the spectral index does not change in the
ProNaOS spectral range. We take into account the variability of the dust
spectral index of the different regions.
This gives us a simple self-consistent
model that allows us to estimate the column density $N_{H}$, as a function of
the spectral intensity and the spectral index.
We adopt the value of the
opacity $\kappa_{100 \mic}$ = 0.361 $cm^{2}/g$ (per gram of total medium: gas
and dust) at 100 $\mic$, because this value is well constrained from IRAS
data.
This is calculated from the extinction curves given in D\'esert \etal
The opacity $\kappa$ is defined as $\tau\over{N_H.m_H}$, thus the relation
between the gas column density and the fit parameters is:

$N_{H} = {C . \lambda^{-\beta}\over\kappa . m_H}$

where $m_H$ is the proton mass and $\kappa$ the gas and dust opacity.

Then for the D\'esert \etal value of $\kappa_{100 \mic}$ we have:

$N_{H}$ = 1.67 $10^{24}$ . C . (100  $\mic)^{-\beta}$

with $N_{H}$ in protons/cm$^{2}$ and C in $\mic^{\beta}$.

We make another simple model using the 100 $\mic$ opacity of the
Ossenkopf \& Henning (1994) model, which is specific to protostellar cores.
The opacity introduced then is $\kappa_{100
\mic}$ = 1 $cm^{2}/g$.
The proportionality relation is:

$N_{H}$ = 6.02 $10^{23}$ . C . (100  $\mic)^{-\beta}$

with $N_{H}$ in protons/cm$^{2}$ and C in $\mic^{\beta}$.

\vspace*{1em}

\subsubsection{Analysis}

We compare our column density results to those derived from observations
of the rotational transition $J = 1-0$ of $^{13}CO$, made by Nagahama \etal
(1998).
We transform the observed $W_{^{13}CO}$ into the column density in
cm$^{-2}$ by using the correspondence established by Bally \etal (1991) in the
Orion Nebula.
This correspondence has been calculated using comparisons between the
IRAS far-infrared dust emission and the CO lines in Orion.
Ristorcelli \etal (1998) have taken for the BN/KL region the coefficient:
$N_{H}/W_{^{13}CO}$ = 2 10$^{21}$ cm$^{-2}$/(K km/s), assuming an excitation
temperature of about 80 K.
We have used the same value of this coefficient for estimating the column
density of OMC-1 from the $^{13}CO$ data.
For the cold and rather cold clouds we study in this article, we choose to
adopt $N_{H}/W_{^{13}CO}$ = 10$^{21}$ cm$^{-2}$/(K km/s), as it is mentioned by
Bally \etal (1991) for an excitation temperature of about 30 K.
Indeed the $^{13}CO$ emission towards the cold clouds we observe ($<$ 20 K) is
not likely to trace very well the clump itself, but rather the surrounding
gas, which has probably a quite higher temperature.
Recent CO observations of Ristorcelli \etal (2001, \inprep) of Cloud 2 seem to
support this, as the $^{13}CO$ emission of this cloud is more diffuse than the
$C^{18}O$ emission.
This method gives a simple way to estimate the gas column density from these
$^{13}CO$ data.
As we can see in Table 5, for OMC-1, ISF-N and ISF-S, there is a rather good
agreement between N$_{H}$ estimated from ProNaOS data by the model from
D\'esert \etal and N$_{H}$ from $^{13}$CO data.
Indeed, the ${N_{H}\tiny{dust}\over N_{H}\tiny{^{13}CO}}$ ratio is 1.3 for
OMC-1, 0.9 for ISF south and 1.5 for ISF north.
Of course the uncertainties on both values are large and difficult to
estimate, but these results prove that these values found are quite robust.
The cold clouds (1,2,3,4) have $^{13}$CO column densities in rather good
agreement with the ProNaOS Ossenkopf \& Henning estimations:
${N_{H}\tiny{dust}\over N_{H}\tiny{^{13}CO}}$ is 0.7 for Cloud 1, 1.2 for
Cloud 2, 1.7 for Cloud 3 and 2.4 for Cloud 4.
This rather good
agreement between the $^{13}CO$ estimated column densities and the O\&H estimated ones may be explained by the fact that some cold condensations are the
site of formation of molecular ice mantles on the grains, and of coagulation
of grains.
These processes are taken into account by the model of Ossenkopf \&
Henning.
However, the $^{13}$CO estimations for N$_{H}$ seem under-estimated for Clouds
3 and 4, compared to the ProNaOS O\&H values (${N_{H}\tiny{dust}\over N_{H}\tiny{^{13}CO}}$ ratios of 1.7 and 2.4).
If we suppose a higher excitation temperature than 30 K for these two clouds,
it is possible to match the ProNaOS O\&H values, for instance assuming
T\ap80 K leads to $10^{21}$ \cm2
for the $^{13}$CO N$_{H}$ value of Cloud
4, but this warm temperature is not likely for this region.
Another possibility may be that the O$\&$H opacity value still over-estimates
the column density for these clouds, that may mean that they are the place of
even more grain special effects than those taken into account by Ossenkopf \& Henning.

\begin{table}
\caption[]{\label{table:interpretation}
Column densities estimated from the opacities of D\'esert \etal (90), Ossenkopf \&
H. (94), and from the $^{13}$CO data of Nagahama \etal (98). Masses, Jeans mass, densities.}

\begin{flushleft}
\begin{tabular}{lllllllll}
\hline

&N$_{H}$ \tiny{D\'esert} & N$_{H}$ \tiny{Ossenkopf} & N$_{H}$ \tiny{$^{13}$CO}
& \tiny{Mass D\'es.} & \tiny{Mass Oss.} & \tiny{Jeans m.} &
\tiny{Density D\'es.} & \tiny{Density Oss.} \\

&$10^{20} cm^{-2}$ &$10^{20} cm^{-2}$ & $10^{20} cm^{-2}$ & (\msol)
& (\msol) & (\msol) & \tiny{(protons/cm${^3}$)} & \tiny{(protons/cm${^3}$)} \\
\hline

OMC-1(3.6') & 1400 & ------ & 1080 & 212 & --- & --- &  138000 & --- \\
\hline

ISF & & & & & & & \\
(south) 5.4' & 245 & ------ & 270 & 83 & --- & --- &  16100 & --- \\
\hline

ISF & & & & & & & \\
(north) 4.2'  & 525 & ------ & 340 & 108 & --- & --- &  44400 & --- \\
\hline

Cloud1 (3.5') & 320 & 115 & 175 & 42 & 15 & 12.6 & 33800 & 12200 \\
\hline

Cloud2 (3.5')  & 320 & 120 & 100 & 46 & 17 & 8.7 & 32900 & 11900 \\ 
\hline

Cloud3 (3.5')  & 95 & 34 & 20 & 9.9 & 3.6 & 8.4 & 11300 & 4070 \\
\hline

Cloud4 (5.2')  & 33 & 12 & 5 & 10.3 & 3.7 & 18.5 & 2220 & 802 \\
\hline

\end{tabular}
\end{flushleft}
\end{table}

We also show in Table 5 the mass of each region. This is obtained simply by
multiplying the estimated gas column density by the surface of the region,
assuming a distance of 470 pc.

For the clouds separated from the molecular complex, we may reasonably
ask if they can collapse due to their mass and therefore become protostellar
objects. As an indicator of the stability of the clouds, we give in Table 5
the Jeans mass. We use for this the
expression derived from the equality of gravitational and thermal energy (\cite{larson69}):

$M_{J}$ = 10$^{-18}$ . ${D_{cm}\over 2}$ . T

with M$_{J}$ in solar masses and $D_{cm}$ the diameter in cm of the cloud.

T refers in this equation to the gas
temperature, whereas we used the dust temperature found.
In the dense
interstellar medium, the two temperatures are weakly different
(\cite{hollenbach88}, \cite{tielens85}).
We assume by this way that the mass is essentially the clump's one, and that
the gas temperature of the clump is about the same as the dust one, but it may
be different from the temperature of the supposed surrounding gas
(traced by the $^{13}$CO).

The comparison of the cloud mass to the Jeans mass is here to be taken
as an indicator of the cloud's instability. There are of course large uncertainties
in both the clump's measured mass and the non thermal sources of internal
energy, such as magnetic field and turbulence, which can help to balance the
gravitational energy.
On the other hand, external pressure may lead to stabilization of otherwise evaporating clumps.
In our sample of four cold clouds, there is a
trend that the clumps closer to the active region (Clouds 1 and 2) are potentially
the more unstable.

\section{Conclusion and summary\label{conc}}

Our study shows a large distribution of temperatures and spectral indexes in
and around a dense and active molecular complex, the M42 Orion
Nebula. The temperature varies from 12 K to 70 K, and the spectral index from 1.1 to
2.2.
The finding of two new cold clouds (Clouds 3 and 4) confirms that the
existence of cold condensations in such regions is not unusual. However the
extended cold
clumps are located in the outskirts of the active star forming area. They may
be the sites for future star formation.

The statistical analysis of the temperature and spectral index spatial
distribution shows an evidence for an inverse correlation between these two
parameters.
This effect is not well explained yet, especially in the submillimeter
spectral range for cold grains ($<$ 20 K). It has been shown to occur in the
laboratory for warm grains
by Mennella \etal (1998), and for cold grains in the millimeter by Agladze
\etal (1996).

We estimated the column densities and masses of the observed regions, by simply
modelling the thermal emission of the grains from D\'esert \etal (1990) or
Ossenkopf \& Henning (1994). There is a good agreement between the $^{13}CO$ column densities and those
derived from our submillimeter measurement. This demonstrates the robustness
of dust opacity values in the grain models.
The submillimeter wide band spectro-imaging is thus a natural way to derive
masses in the interstellar medium.

Finally we see a trend that the closer to the
complex the cold clouds are, the more unstable they are. The history of star formation around OMC-1 shows
that there have already been 3 to 4 successive bursts of star formation
in this region with the embedded cluster responsible for the BN/KL object
being the latest. The clouds that we observe close to the active region may
thus be the seeds of the next generation of stars. This is of course quite
speculative and should be sustained by more observations, particularly of the
possible embedded protostars.

\section{Acknowledgements}

We wish to thank Drs Nagahama, Mizuno, Ogawa and Fukui, for kindly having
provided us their CO maps.
We are indebted to the French space agency Centre National d'\'Etudes Spatiales
(CNES), which supported the ProNaOS project. We are very grateful to the
ProNaOS technical teams at the CNRS and CNES, and to the NASA-NSBF balloon-launching facilities group of Fort Sumner (New Mexico).

\end{document}